\documentclass{llncs}
\usepackage{llncsdoc}
\usepackage[table,xcdraw]{xcolor}
\usepackage{graphicx}
\usepackage{booktabs}
\usepackage{adjustbox}
\usepackage{multirow}
\usepackage{diagbox}
\usepackage{tablefootnote}

\begin{document}
\title{Spotting Information biases in Chinese and Western Media}
\author{Dominik Wurzer \inst{1}  \and Yumeng Qin \inst{2} }
\institute{$^1$ School of Informatics, University of Edinburgh, UK\\
$^2$ International School of Software, Wuhan University, China}

\maketitle

\begin{abstract}
Newswire and Social Media are the major sources of information in our time. While the topical demographic of Western Media was subjects of studies in the past, less is known about Chinese Media. In this paper, we apply event detection and tracking technology to examine the information overlap and differences between Chinese and Western - Traditional Media and Social Media. Our experiments reveal a biased interest of China towards the West, which becomes particularly apparent when comparing the interest in celebrities.
\end{abstract}
\section{Introduction}
Historically information was reserved for a minority, who controlled it. The emergence of the Internet and Social Media services, which provided free access of information to everyone, acted as an equalizer. Researchers, like [3] and  [7], identified Traditional Media (TM) - in the form of news-wire articles and Social Media (SM) - including Twitter, Facebook and Blogs, as the two main information sources of our time. [5] and [4] studied the difference in information provided by Western TM and SM by quantitatively determining the information overlap between them. They found that both, TM and SM, provide broad coverage of major news topics, while SM additionally carries minor events that were ignored by TM. All studies to this data compare information overlap of Western (European and North American) TM and SM. In this paper we address two main research questions: \textbf{(1)} Do studies of western TM and SM also apply to Chinese TM and SM? and \textbf{(2)} What are the differences in information shared by Western and Chinese TM and SM?\\\\We argue that it is interesting and important to study the topical demographic of media in China, since it is disjointed with the rest of the world. In China, access to foreign TM and SM web-sites is restricted and Chinese TM is controlled by the government. For example, Facebook, Twitter and the Wall Street Journal\footnote{http://www.wsj.com/} are not reachable in China. Instead, Chines citizens use equivalent services offered by Chinese companies. This raises questions about which information is shared by both and which is not.
\begin{table*}[h]
\centering
\begin{tabular}{|c|c|c|c|}
\hline
Name & Source &  Documents   & Detected \\ 
 & &  &  Topics \\ \hline
Western Media & CNN, BBC, New York Times, Google News & 60k &  42k \\ \hline
Western & Twitter\tablefootnote{Twitter Streaming API https://dev.twitter.com/streaming/} & 50 mio &  2.1 mio\\\cline{2-4}
Social Media & Facebook\tablefootnote{Facebook Search Graph} & 8 MIO & 230K \\ \hline
Chinese Media & Xinhua,CCTV-News,Sohu, Baidu News & 55k &  37k \\ \hline
Chinese & Sina Weibo\tablefootnote{Weibo API http://open.weibo.com/} & 50 mio &  2.3 mio\\ \cline{2-4}
Social Midea & RenRen\tablefootnote{RenRen API http://dev.renren.com} & 9 mio &  257k\\ \hline
\end{tabular}
\caption{Data set statistics of Chinese and Western SM and TM}
\label{tab:data}
\end{table*}
\vspace{-2mm}\section{Data Set}  
\vspace{-1mm} We compare two types of media: SM and TM. Each type of media is represented by major corporations in China and the Europe/USA, -dubbed Western, as seen in Table \ref{tab:data}. Sina Weibo is the equivalent of Twitter in China and RenRen is comparable to Facebook. Table \ref{tab:data} shows the number of documents we crawled during 76 days days from the period of June $1^{st}$ 2016 to August $15^{st}$ 2016. The number of detected topic results from an automated state-of-the-art topic detection, discussed in Section \ref{sec:Methodology}. 

\section{Methodology}
\label{sec:Methodology}
\begin{figure*}[h]
\centering
\includegraphics[width=\textwidth]{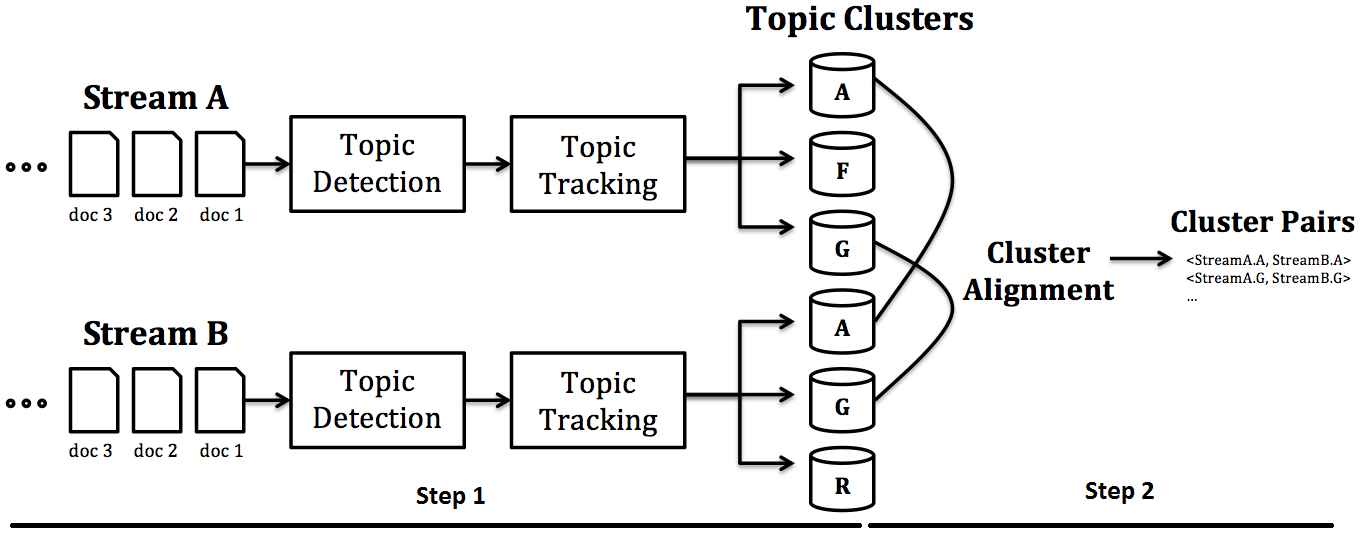}
\caption{illustrating the methodology for computing the overlap of topics discussed in two streams}
\label{fig:blockgraph}
\end{figure*}

\vspace{-1mm}Our approach to determining the difference in information overlap between two streams is twofold, as in Figure \ref{fig:blockgraph}. 
In a first step, we identify topics using k-term hashing, a state-of-the-art topic detection technology [6]. We apply k-term hashing to each of the 4 data streams (Western and Chinese, TM and SM). The set of identified topics is fed into an adjacent Topic Tracking system, where they are used to initialize the topic clusters. Our tracking system - adjusted for a high precision setting - builds clusters by grouping documents of the same topicality using \textit{tf.idf} weighted cosine similarity. Each cluster is represented by a centroid vector, whose feature count is limited to k and computed based on the average term statistics of its associated documents. Since SM streams are highly noisy, we remove singleton clusters and apply standard IR preprocessing methods. This leaves us with a set of topic clusters for each stream, as in Figure \ref{fig:blockgraph}. In a second step, we determine the degree of overlap between the 2 streams. We align the streams using nearest-neighbour-search based on the proximity of their centroid vectors in term space, measured by a $tf.idf$ weighted dot product. To ensure high-precision, we constrain nearest-neighbour-search to only align two topics when their pair-wise similarity is high.
\begin{table}[h]
\centering
\begin{tabular}{|c|c|c|c|c|c|c|}
\hline
\diagbox{from}{to} & CHN TM & RenRen & Weibo & Western TM&  Twitter & Facebook  \\  \hline
CHN TM & - & 68\% & 98\% &  43\% & 23\% & 9\% \\ \hline
RenRen & 11\% & - & 16\% & $<$1\% & $<$1\% & $<$1\% \\ \hline
Weibo & 8\% &  6\%& - & 4\% & 9\%  & 3\% \\ \hline
Western TM& 42\%  & 16\% & 38\% & - & 99\% & 86\%  \\ \hline
Twitter & 4\% & $<$1\% & 9\% & 8\% & - & 7\% \\ \hline
Facebook & 1\% & $<$1\% & 3\% & 9\% & 14\% & -  \\ \hline
\end{tabular}
\caption{Information overlap between Chinese and Western TM and SM}
\label{tab:overlap}
\end{table}

The degree of information overlap equates to the percentage of topics that were successfully aligned between two streams. Table \ref{tab:overlap} provides the degree of information overlap between all data streams. Note that stream alignment is a directed a-symmetric relation ship, thus $A \rightarrow B \neq B \rightarrow A$. 
\section{Determining the Coverage of Chinese Traditional Media and Social Media}
\label{sec:ChineseTraditionalMediaAndSocialMedia}
Our first experiment determines the coverage of Chinese TM by its SM and vice-versa.\\\\
\textbf{Chinese Traditional Media vs. Social Media}\\ Table \ref{tab:overlap} reveals a substantial difference in the coverage of Chinese TM topics between Sina Weibo, which covers 98\% and RenRen, which only covers 68\%. We further inspect, which topics are not shared by $TM\rightarrow SM$ and sampled 100 TM topics that were not found in the SM stream. Out 100, we identified 16 topics that were in fact covered but not recognized by our stream alignment method. The remaining 84 topics include minor foreign politics topics, like the visit of a Vice President, financial news of small based Singaporean and US companies and sport related news, like the comments of an IOC member. We conclude that Chinese Social Media services and Weibo in particular, have a very good coverage of Chinese newswires.\\\\
\textbf{Chinese Social Media vs. Traditional Media}\\SM streams produce a far greater number of topics than TM streams. Not all of these topics are news worthy. In our next experiment we determine, whether Chinese SM contains relevant information, not reported by Chinese TM. We investigate unaligned topics and randomly sampled 3,000  for manual inspection. \textit{Note}: these are topics discussed on SM, for which no similar topic was found in the TM stream. Out of these 3,000 topics, 2,356 contained trivial chatter and 644 reported about actual events covering: celebrity news (227), minor accidents (139), information on public transport and street closures (112), sport events (42), the opening of restaurants, boutiques and hotels (34) a mix of different kind of events (110).\\\\
We conclude that the vast majority of topics discussed by the official Chinese TM are also actively discussed on Chinese SM. SM provides additional information about celebrities, local events, as well as results of minor sports events. These findings overlap with [5] and answer our first reach question: The relationship between Chinese TM and SM is comparable to the relationship of Western TM and SM.
\subsection{Interpreting Alignment Intensity}
In addition to coverage, we are interested in the intensity, with which SM streams overlap with TM streams. For example, \textit{which news topics trigger the most discussions on SM?} We define 4 topic categories: celebrity, political, accidents/disasters and financial news. Luckily, about 20\% of the articles in our TM data set come with category labels, which we harness as training data. Using a classifier, based on language models in conjunction with content features, we assign TM topics to the 4 categories and select the 500 highest ranked topics for each of them. Manual verification ensures that the 2,000 topics are correctly classified. Coverage intensity is measured by the number of SM messages aligned to the 500 TM topics in each topic category. Celebrity related news receive by far the most attention from SM users. This is interesting, since celebrity news only make up less than 10\% of all TM topics. The second most attention receive topics covering accidents and disasters, followed by financial news. Interestingly, political news seems to be less actively discussed on Chinese SM. 
We conjecture that the reduced intensity of political topics could be linked to censorship or general posting restrictions. For example, critical posts about politicians, riots or protest movements are likely to censored and vanish from Sina Weibo [1]. 
\section{Information Coverage of Chinese and Western Traditional Media}
Before we can align Chinese and Western TM, we translate all text written in Mandarin to English. We apply Moses [2], a phrase-based statistical machine translation system that has been trained on a newswire corpora and is known for its high translation quality. \\\\Our initial assumption was that both streams report about international topics, in addition to domestic topic that are unique to them. Following our stream alignment we measure a topic overlap of 43\%. We are curious whether the coverage is biased and randomly sample 3,000 aligned topics. We categorize them on whether they are related to international or domestic topics. Out of the 3,000 documents, 2,152 (72\%) covered international events, like Olympics and acts of terror carried out by ISIL. Interestingly, 709 (23\%) topics describe events in Europe and the USA like, USA extends sanctions on Russia and a German minister's report on crime committed by refugees. By contrast, only 139 (4\%) discussed events that took place in China, which included financial news like an enterprise reform symposium in Beijing, or political news covering the South Chinese See and the crackdown on pro-democracy protesters. By contrast, Chinese TM appears to cover western events of smaller granularity and in more detail. \\\\We further inspect 200 articles from Western TM that report about China, for which we could \textit{not} find a similar article of a Chinese TM. Out of these 200 news articles, 71 covered news about politicians and the communist party, 64 reported about environmental issues, 31 talked about foreign affairs including China South Sea and Taiwan, 22 reported about dissidents and artists, 12 had in fact a Chinese article that was not correctly aligned. We conjecture that these articles represent \textit{sensitive} topics for the state-controlled Chinese TM. 
\section{Aligning Chinese and Western Social Media}
Before aligning Western and Chinese SM, we translate all text written in Mandarin to English. The mutual coverage of both streams is rather low in comparison with TM. To gain insight into what information is covered by both streams, we apply k-mean clustering. In particular, we define 30 random seeds and cluster to all topics that appear in both streams. The resulting 30 clusters are manually examined to determine what type of topics they incorporate. By far the biggest cluster discusses celebrity related information, followed by discussion of gadgets, consumer goods and brands.\\\\
 \subsection{Biased Interest in Celebrities}
The previous section revealed that Western and Chinese SM actively discuss celebrity related news. We are interested, whether the interest in celebrities is biased and extract 500 persons born in the USA and 500 persons born in China from Wikipedia. We limit the age range from 20 to 35 and only target persons that are currently active as musicians, actor or athletes. We then apply rule based named entity recognition, using name variations and abbreviations extracted from Wikipedia, to measure the number of mentions in Western and Chinese SM. Interestingly, we found a high bias towards celebrities born in the USA. Nearly all (88\%) of the celebrities mention on Western SM and half (48\%) of the celebrities mentioned on Chinese SM, are born in the USA. By contrast Chinese born celebrities only receive 12\% of the mentions in Western SM and 52\% in Chinese SM. We conclude that the interest in celebrities is highly biased. The users of Western SM highly favour celebrities born in the USA. By contrast, users of Chinese SM showed a more balanced interest in celebrities.


\vspace{-0.5mm}\section{Conclusion}
\vspace{-1.5mm}In this paper we studied the information overlap of Chinese and Western Traditional Media (TM) and Social Media (SM). Our study suggests that Chinese SM covers most of the topics discussed by Chinese TM and provides additional information about celebrities and locally relevant events. When comparing Western and Chinese TM and SM, we found a bias of China towards the West, as Chinese TM reports small scaled western events, while 
 Western TM mainly focuses only on major news about China. This trend becomes particularly apparent when comparing the interest of SM users in celebrities. While Western SM users barely show any interest in Chinese born celebrities, Chinese SM users actively discuss both Chinese and Western born celebrities. We also revealed several \textit{sensitive} topics reported by Western TM, for which we could not find a corresponding Chinese article. We assume that these contain unfavourable information from the point of view of the state controlled Chinese TM.


\begin{thebibliography}{}
1. Gary King, Jennifer Pan, and Margaret E Roberts. Reverse-engineering censorship in china: Randomized experimentation and participant observation. 2014. \\
2. Philipp Koehn, Hieu Hoang, Alexandra Birch, et al. Moses: Open source toolkit for statistical machine translation. In Proceedings of the 45th annual meeting of the ACL on interactive poster and demonstration sessions, pages 177–180. ACL, 2007.\\
3. Haewoon Kwak, Changhyun Lee, Hosung Park, and Sue Moon. What is twitter, a social network or a news media? In Proceedings of the 19th international conference on World wide web, pages 591–600, 2010.\\
4. Miles Osborne and Mark Dredze. Facebook, twitter and google plus for breaking news: Is there a winner? In ICWSM, 2014.\\
5. Sasa Petrovic, Miles Osborne, Richard McCreadie, Craig Macdonald, and Iadh Ounis. Can twitter replace newswire for breaking news? 2013. \\
6. Dominik Wurzer, Victor Lavrenko, and Miles Osborne. Twitter-scale new event detection via k-term hashing. In Proceedings of the 2015 Conference on EMNLP, pages 2584–2589, Lisbon, Portugal, September 2015.\\
7. Wayne Xin Zhao, Jing Jiang, et al. Comparing twitter and traditional media using topic models. In European Conference on Information Retrieval, pages 338–349, 2011.
 
%
\end{thebibliography}
\end{document}